\documentclass[a4,11pt]{cip-v3-cls-submit} 
\usepackage{afterpage}
\usepackage{changes}

\usepackage{mathptmx,amsmath}
\usepackage{hyperref, cases}

\hypersetup{colorlinks=true, citecolor={blue}, linkcolor= {blue}, urlcolor ={blue},	pdfborder={0 0 0}
}

\def\Mr{\uppercase}

\usepackage{float}
\usepackage{amsmath}

\usepackage{subfigure}

\usepackage{wasysym}
\usepackage{graphicx,wrapfig,tikz}
\usepackage{caption} 
\usepackage{amsmath,amsxtra,amssymb,latexsym,amscd,color,bm}
\usepackage[mathscr]{eucal}
\usepackage{epstopdf}
\usepackage{array} 
\usepackage{cite}

\def\vsm{\vskip0.1cm}

\def\doi#1{\href{http://dx.doi.org/#1}{doi:#1}}

\def\titles#1{\title{\large\bf\noindent #1}}
\def\authors#1{\author{\begin{flushleft}{#1}\end{flushleft}}}
\def\authord#1#2{\indent\Mr{#1}\\
	\textit{\indent#2}\vsm}

\def\email#1{\bigskip\href{mailto:#1}{\textit{E-mail:}~{#1}}\\[3mm]}

\def\received#1{\vsm\textit{\indent Received #1}}
\def\accepted#1{\vsm\textit{Accepted for publication~#1}}

\def\Keywords#1{\\[.2cm] Keywords:~{#1}.}

\def\and{$\text{\tiny AND }$}

\def\Classification#1{\\[.2cm] Classification numbers:~{#1}.}

\begin{document}
\Year{2025}
\Page{1}\Endpage{8}
\titles{Effect of linear and quadratic coupling on dynamical parameters of an optomechanical oscillator}

\authors{	\authord{Le Tri Dat$^{\lowercase{a}}$, Vinh N.T. Pham$^{\lowercase{b}}$, Le Van Tan$^{\lowercase{c,d}}$, and Nguyen Duy Vy$^{\lowercase{c,d},*}$}{
$^a$Nuclear Training Center, VINATOM, 140 Nguyen Tuan Str., 10000 Ha Noi, Vietnam\\
$^b$Department of Physics \& Postgraduate Studies Office, Ho Chi Minh City University of Education, Vietnam\\
$^c$Laboratory of Applied Physics, Science and Technology Advanced Institute, Van Lang University, Ho Chi Minh City, Vietnam\\
$^d$Faculty of Applied Technology, School of Technology, Van Lang University, Ho Chi Minh City, Vietnam}
$^*$\email{nguyenduyvy@vlu.edu.vn}
\received{\today}
\accepted{DD MM 2025}	}

\maketitle

\markboth{L.T. Dat, V. Lan, and N.D. Vy}{Exact mode shapes of T-shaped and overhang-shaped microcantilevers}

\begin{abstract}
Dynamics of icrocantilevers are of important interest in micro-mechanical systems for enhancing the functionality and applicable range of the cantilevers in vibration transducing and highly sensitive measurement. In this study, using the semi-classical Hamiltonian formalism, we study in detail the modification of the mechanical frequency and damping rate taking into account both the linear and quadratic coupling between the mechanical oscillator and the laser field in an opto-mechanical system. It has been seen that, the linear coupling greatly enhances the modification of the effective mechanical frequency and the effective damping rate while the quadratic coupling reduces these quantities. For a MHz-frequency oscillator, the damping rate could be 10$^5$ times increased and the frequency is several times modified. These results help clarifying the origin of the modification of the susceptibility function for cooling of the mechanical mode.
\Keywords{microcantilever, mode shape, analytical method, overhang-shaped, T-shaped}
\Classification{07.79.Lh, 78.20.N-, 65.40.De}
\end{abstract} 

\section{Introduction}
Light-matter coupling at macroscopic to mesoscopic scales has garnered significant interest over the past two decades to elucidating quantum-to-classical physics. 
In the experiment for seeking the gravitational wave from the universe, a test mass can be controlled to reveal the extremely subtle impact of gravitational waves on Earth \cite{16LaskyPRL_LIGO, 20Mciver, 19TsePRL_LIGO, 19JungPRL_LIGO}. In the expariment for revealing the quantum-to-classical transition, the oscillators of pico-gram in mass were used and its effective temperature could reduced to the millikelvin range, where the phonon number approaches zero \cite{14MeenehanPRL_cool, 16PetersonPRL_cool, 16HarrisNatPhys_cool, 20QuiPRL_cool}. 
In such systems, reducing the noise from environmental sources, the quantum properties of the driving light, or the mutual interaction between light and object is crucial. A system comprising two mirrors that trap and amplify an electromagnetic field (a laser) has proven highly effective in this domain. When the distance between the two mirrors, the cavity length $L_c$ is close to a multiple of the laser wavelength, a small change in the cavity length reduces the stored field and the radiation force exerted on the mirror. As a result, the coupling between the mechanical and optical modes is maintained. 
%



Beside the setting where the linear optomechanical coupling is available, the system could be created where the second-order coupling is seen, for example, membranes are inserted inside an optical cavity by Harris et al. \cite{08HarrisNJP_middle,16HarrisNatPhys_cool}, by Vitali et. al. \cite{VitaliPRLmirrorField},  Purdy et al. \cite{13PurdyPRX}, Favero et al. \cite{FaveroAPLcool07,09FaveroOE} and Weig et al. \cite{13WeigAPL}. A detailed discussion of these system could be seen by Favero et al. \cite{09FaveroOE}. And the quadratic coupling give fruitful contribution on the cooling in optomechanics.

In this study, we examine the contribution of both the linear and the quadratic coupling on the modification of the mechanical frequency and the dampind rate of the oscillator in an optomechanical system. These two parameters are crucial in controlling of the dynamics of the oscillator and any change of them could lead direatly to the modification of the susceptibility function, the function that determines the effective temparature and oscillating (mechanical) energy. Therefore, a detailed study of the modification of the frequency and damping rate is of inerest.
\begin{figure}[!h] \centering
\includegraphics[width=1.0\textwidth]{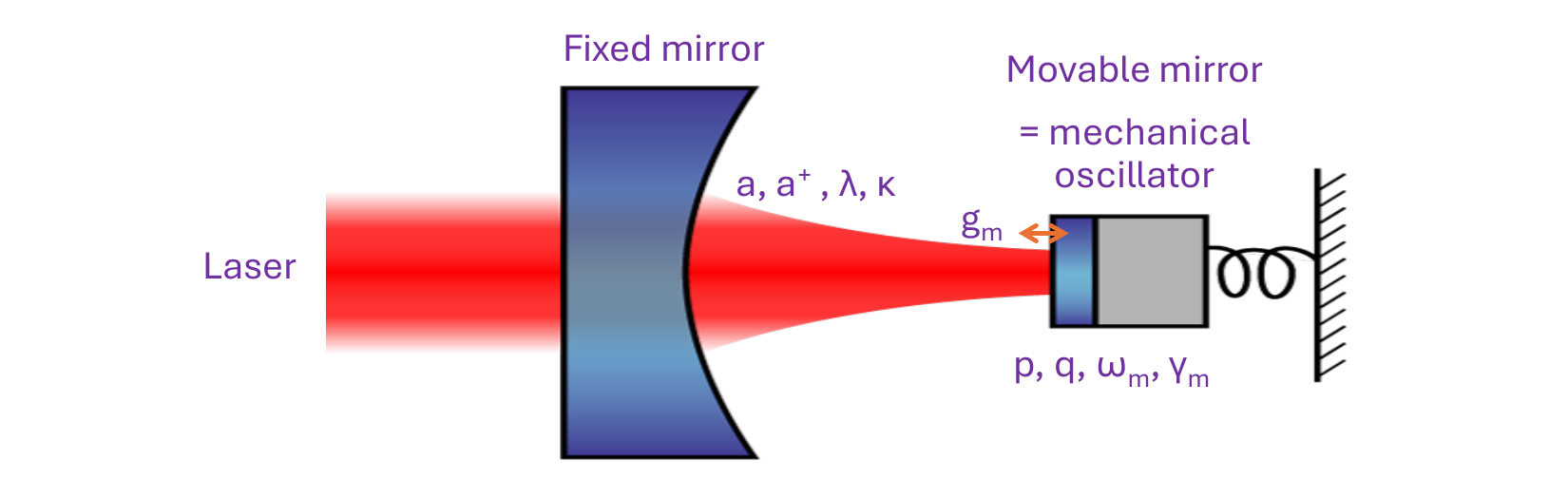}
\caption{(top) Photon $S_a$ and (bottom) phonon $S_q$ spectra for increasing opto-mechanical coupling strength $G$. A reduction in $S_q$ implies a cooling of the mechanical oscillator, and this is corresponding to an increase in $S_a$, i.e. the energy of thermal bath has been transferred from the mechanical to the optical mode.} \label{fig_1}
\end{figure}

The optomechanical system involving an optical microcavity irradiated by a laser is shown in Fig. \ref{fig_1}. 
The first mirror is a semitransparent thin film, and the second mirror plays the role of a mechanical oscillator, which is movable and has a mechanical frequency of $\omega_m$, a decay of $\gamma_m$, and located at $L_C$ from the first mirror. The thicknesses of the two films were chosen so that they could maximize the effect of the radiation pressure on the two inner surfaces \cite{VyAPEX16, 17Hoang_OC,Vyapex20,22Dat_OC}.

\section{Hamiltonian formalism for optomechanical oscillators}
The coupling between the optical and the mechanical mode is assumed to involve the linear and quadratic \cite{13Xuereb_quad, 22Banerjee_quadratic, 24Ghorbani_quadratic} terms as 
\begin{align}
H_i= -\hbar g_ma^\dagger a q +\hbar g_2a^\dagger a q^2, 
\end{align}
where 
$g_m =\frac{\partial\omega_c}{\partial x}x_{ZPF} =\frac{\partial\omega_c}{\partial x}\sqrt{\frac{\hbar}{2\omega_m m}}
$ 
is the coupling strength, and
$\frac{\partial \omega_c}{\partial x} \simeq \frac{\omega_c}{L_c}, $
which leads to $g_m =\frac{\omega_c}{L_c} x_{ZPF}$. 
As a result, we obtain the total Hamiltonian (in the rotating frame of the pump laser with frequency $\omega_p$) where the laser has a power of $P_i$, an amplitude of $\epsilon_p=[P_i\kappa/(2\hbar\omega_p)]^{1/2}$, and a cavity damping rate of 2$\kappa$
\begin{align}
H=&\hbar\Delta_0 a^\dagger a +\frac{1}{2}\hbar\omega_m(p^2+q^2) -\hbar g_ma^\dagger aq  +\hbar g_2^\dagger aq^2 +
i\hbar\epsilon_p(a^\dagger-a), \label{Htot}
\end{align}
 where $\Delta_0 = \omega_c-\omega_p$ and $\omega_c=2\pi c/L_c$ is the frequency of the photon mode inside the cavity. 
Using the Heisenberg equations for the operators $a$, $a^\dagger$, $q$, and $p$, such as $-i\hbar \dot{a}=[a,H]$, 
\begin{subequations} \label{eq.Langevin}
\begin{align}
\dot{a}&= -i\Delta_0a +i g_m aq -i g_2 a q^2 +\epsilon_p, \label{dot_a_}\\
\dot{a}^\dagger &=  i\Delta_0 a^\dagger -i g_m a^\dagger q +i g_2 a^\dagger q^2  +\epsilon_p, \label{dot_adag_}\\
\dot{p}&=  -\omega_m q +g_ma^\dagger a -2g_2 a^\dagger a q , \label{dot_p_}\\
\dot{q}&= \omega_mp,
\end{align}\end{subequations}
then adding the terms of noises and the dempings, we obtained the following equations,
\begin{subequations} \label{eq.Langevin}
\begin{align}
\dot{a}&= -(\kappa+i\Delta_0)a +i g_m aq -i g_2 a q^2 +\epsilon_p +\sqrt{2\kappa}a_{in}, \label{dot_a}\\
\dot{a}^\dagger &= -(\kappa-i\Delta_0) a^\dagger -i g_m a^\dagger q +i g_2 a^\dagger q^2  +\epsilon_p +\sqrt{2\kappa}a_{in}^\dagger, \label{dot_adag}\\
\dot{p}&= -\gamma_m p -\omega_m q +g_ma^\dagger a -2g_2 a^\dagger a q +\xi(t), \label{dot_p}\\
\dot{q}&= \omega_mp.
\end{align}\end{subequations}
The cavity mode is affected by the input noise from the vacuum radiation $a_{in}$ and the mechanical mode is affected by the fluatuation $\xi$. They follow the correlation functions \cite{walls2008quantum,HuangNJP09,GioavaPRA01noise, NewJPdyn} 
 \begin{subequations}\label{corre.phot}\begin{align}
\langle a_{in}(t)a_{in}^\dagger(t')\rangle &=[N(\omega_c)+1]\delta(t-t'),\\
\langle a_{in}^\dagger(t)a_{in}(t')\rangle &=N(\omega_c)\delta(t-t'),\\
\langle a_{in}(t) a_{in}(t')\rangle 
&=\langle a_{in}^\dagger(t) a_{in}^\dagger(t')\rangle =0,
\\
\langle \xi(\omega)\xi(\omega')\rangle & =\frac{\gamma_m}{\omega_m}\omega\big[\coth(\frac{\hbar\omega}{2k_BT})+1\big]\delta(\omega+\omega'), 
\end{align}
\end{subequations}
where 
$N$ 
is the two-photon correlation functions \cite{HuangNJP09}. 
The steady state solutions are obtained by setting the derivatives to zero and we have
\begin{eqnarray} \label{steady}
\begin{cases} a_s =& \dfrac{\epsilon_p}{\kappa +i(\Delta_0 -g_mq_s +g_2q_s^2) } \\
a_s^\dagger =& \dfrac{\epsilon_p}{\kappa -i(\Delta_0 -g_mq_s+g_2q_s^2) }\\
q_s =& g_m|a_s|^2/(\omega_m +2g_2|a_s|^2) \\
p_s =&0.\end{cases}
\end{eqnarray} 
To obtain the fluctuation spectra of the transmitted field, we linearize the quantum Langevin equation by writing the operators as the summation of their mean values, and the fluctuation operators \cite{VitaliA84phasenoise}, such as $a=a_s+\delta a$. We then keep the linear terams and skip all terms that is higher than 2nd order of fluctuations, such as $\delta a^2$ or $\delta a \delta q$. Thus, we obtain,
\begin{subequations}\label{dot.equ1}\begin{align}
\delta\dot{a} 
 =& -(\kappa+i\Delta)\delta a +iG_a\delta q +\sqrt{2\kappa}a_{in},\\
\delta \dot{a}^\dagger =& -(\kappa-i\Delta)\delta a^\dagger -iG_a^*\delta q +\sqrt{2\kappa}a_{in}^\dagger,\\
\delta\dot{p} 
 =& -\gamma_m\delta p -(\omega_m+2g_2|a_s|^2)\delta q +G_a^*\delta a+G_a\delta a^\dagger +\xi ,\\
\delta\dot{q}=& \omega_m\delta p,
\end{align}\end{subequations} where $\Delta=\Delta_0-g_mq_s +g_2q_s^2$ and $G_a = (g_m -2g_2q_s) a_s$. 
Taking the Fourier transform  $\mathcal{F}[\delta \dot{a}(t)] \rightarrow -i\omega\delta a(\omega)$, Eq. (\ref{dot.equ1}) could be rewritten in a matrix form as follow, 
\begin{align}
 \begin{pmatrix}
 -i\omega+\kappa+i\Delta & 0 & 0 & -iG_a \\ 
 0 & -i\omega+\kappa -i\Delta & 0 & iG_a^* \\ 
 - G_a^* & - G_a & -i\omega+\gamma_m & \omega_m+2g_2|a_s|^2 \\ 
 0&0& -\omega_m & -i\omega 
 \end{pmatrix} \left( \begin{array}{c} \delta a \\ \delta a^\dagger \\\delta p \\\delta q \end{array} \right) = 
 \left( \begin{array}{c} \sqrt{2\kappa}a_{in} \\ \sqrt{2\kappa}a_{in}^\dagger \\ \xi \\ 0 \end{array} \right). \label{eqFourier}
\end{align} 
Assuming that the Routh-Hurwitz criterion for the parameters is satisfied, 
then Eq. (\ref{eqFourier}) has solutions \cite{VyAPEX15}. 
From Eq. (\ref{steady}), we could choose the relative phase reference for the intracavity field and the external laser so that $a_s$ is real and positive, for example, $$\epsilon_p =|\epsilon|e^{-i\theta} =|\epsilon|\frac{\kappa+i(\Delta_0 -g_mq_s+g_2q_s^2)}{\sqrt{\kappa^2+(\Delta_0 -g_mq_s+g_2q_s^2)^2}},$$ we denote $G^*_a = G_a =G$. 
The solution of Eq. (\ref{eqFourier}) is
\begin{align}
\delta q(\omega) =& \frac{-\omega_m}{d(\omega)}
\Big\{
\big[\Delta^2+(\kappa-i\omega)^2\big] \xi 
-i G\sqrt{2\kappa}\big[(\omega+i\kappa-\Delta)a_{in}^\dagger +(\omega+i\kappa+\Delta)a_{in} \big]
\Big\}, \label{delq}
\\
\delta p(\omega) =& \frac{-i\omega}{\omega_m} \delta q(\omega), \label{delp}
\end{align}
and $\delta a$ and $\delta a^\dagger$ that are not shown here for brevity and because we are concentrating on the effect of the opto-mechanical coupling on the phonon variance only.
In Eqs. (\ref{delq}) and (\ref{delp}), 
\begin{align}
d(\omega) =2\Delta G^2\omega_m +[(\omega+i\kappa)^2-\Delta^2][\omega_m(\omega_m+g_2)-\omega^2-i\omega\gamma_m]. \label{eq14}
\end{align}
Dividing the denominator and numerator on the rhs. of Eq. (\ref{eq14}) to ${\Theta=(\omega+i\kappa-\Delta)(\omega+i\kappa+\Delta)}$ and letting
let $\frac{G^2\omega_m }{\Theta}=\Xi$, we obtain
\begin{align}
d_\Theta(\omega) = & d(\omega)/\Theta = \omega_m^2-\omega^2-i\omega\gamma_m +2\Delta \Xi 
= \omega_m^2-\omega^2-i\omega\gamma_m +2\Delta (Re[\Xi]+ i.Im[\Xi])\nonumber\\
= & 
\omega_{eff}^2-\omega^2-i\omega\gamma_{eff}.
\end{align}
Finally, the effective mechanical frequency and the effective damping are written as
\begin{subequations}\label{omgam.effGe}
\begin{align}
\omega_{eff}^2(\omega) &= \omega_m^2 +2\Delta.Re[\Xi]
= \omega_m^2+G^2\omega_m\frac{2\Delta (\omega^2-\Delta^2-\kappa^2)}{[(\omega-\Delta)^2+\kappa^2](\omega+\Delta)^2+\kappa^2]},\label{om.effGe}\\
\gamma_{eff}(\omega) & = \gamma_m +2\Delta.Im[\Xi]
= \gamma_m+G^2\omega_m\kappa\frac{4\Delta}{[(\omega-\Delta)^2+\kappa^2](\omega+\Delta)^2+\kappa^2]}. \label{gam.effGe}
\end{align}
\end{subequations}

We have obtained the analytical formula for the mechanical frequency and the damping rate of the oscillator, which depend on the coupling strength $G$, the de-tuning $\Delta$, and the linear and quadratic couplings.

\section{Results and Discussion}
To present the results, we used these parameters: $\omega_m$ = 2$\pi\times$10$^6$ Hz, $\gamma_m$ = 2$\pi\times$260 Hz, $m$ = 5 ng. The laser to detune and drive the mechanical oscillator has a wavelength of $\lambda$ = 1064 nm, a decay rate $\kappa$ = 6$\pi$$\times{10}^6$ Hz, and an input power $P$ = 0.1--10 mW.

\begin{figure}[!h] \centering
\includegraphics[width=.99\textwidth]{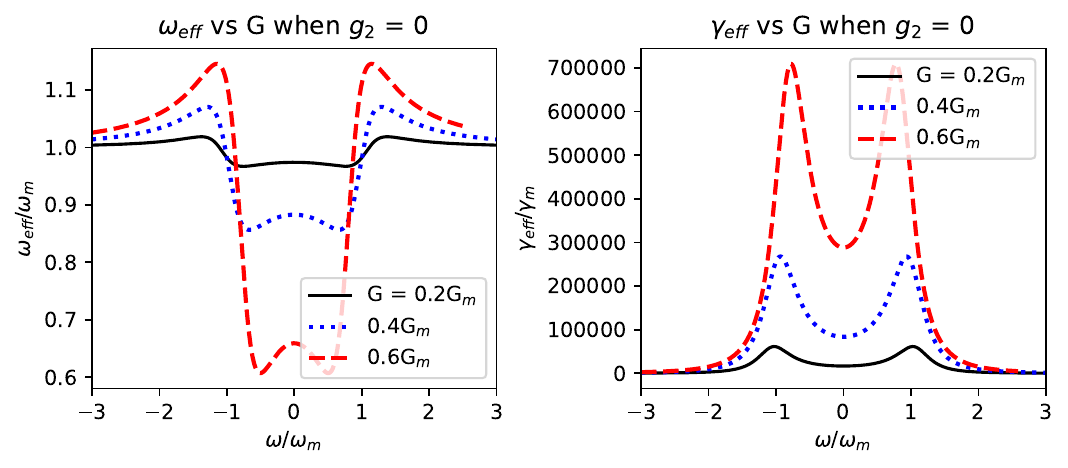}
\caption{(left) Normalized effective mechanical frequency $\omega_{eff}$ and (right) effective damping rate $\gamma_{eff}$ versus the optomechanical coupling strength $G$. Great modifications of frequency and damping lead to changes in the suscapeibility function $\chi(\omega)$ [see. Eq. (\ref{eq_chi})] and this is the mechanism of the passive laser cooling.} \label{fig_gm}
\end{figure}

In Fig. \ref{fig_gm}(left), we present the change of the effective mechanical frequency versus the optomechanical coupling strength $G$. We could see that $\omega_{eff}$ is significantly changed versus $G$. For increasing $G$, $\omega_{eff}$ increases for $\omega>$ 1 and decreases for  $\omega<$ 1. At $\omega=\omega_m$, $\omega_{eff}$ = $\omega_m$. It is worth noting that these modifications will directly to change in the susceptibility function $\chi(\omega)$ of the mechanical oscillator under the exerting of external noise, e.g. from the thermal noise. As we known, from the equation of motion of the oscillator, the variance of the position could be expressed as
\begin{align}
    \langle x^2(\omega) \rangle = \chi(\omega) F_{noise}(\omega), \label{eq_chi}
\end{align}
where $F_{noise}(\omega)$ here denotes the total effects of external noise from both the photonic and bosonic modes. As a result, the oscillation amplitude is modified. If one reduces the value of the function $\chi(\omega)$, the oscillator will reduce the effects from the environment, and could be significantly cooled. Karrai et al. \cite{KarraiNat} used such a mechanism to cool a microcantilever from the room temperature down to 18 K using a milliwatt-laser source. In the right panel, the effective mechanical damping $\gamma_{eff}$ is shown. $\gamma_{eff}$ here is significantly increased, upto 10$^5$ times the original damping rate of several tens of Hz. Accompanying with the effective mechanical frequency, this enhanced damping contributes to the modification of the susceptibility function.

\begin{figure}[!h] \centering
\includegraphics[width=.99\textwidth]{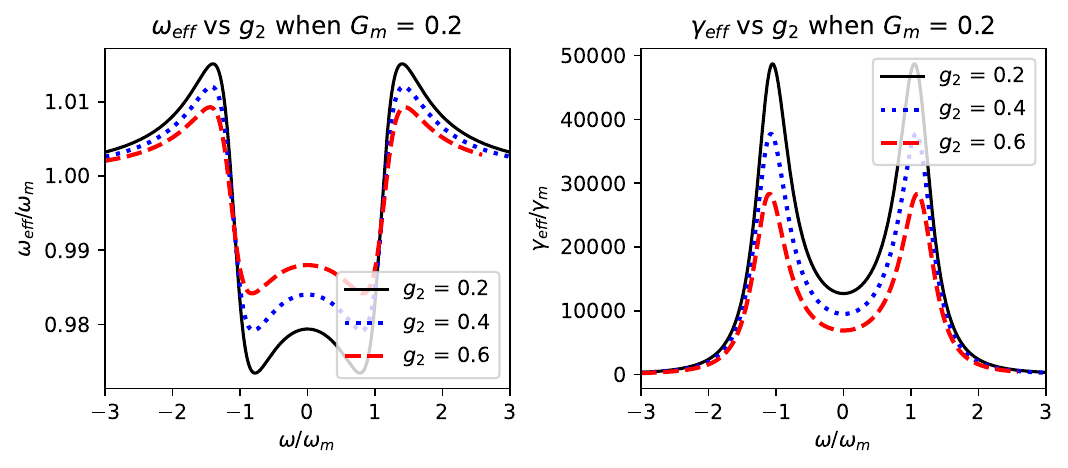}
\caption{(left) Normalized effective mechanical frequency $\omega_{eff}$ and (right) effective damping rate $\gamma_{eff}$ versus the quadratic coupling strength $g_2$ for a fixed value of linear coupling $G$. The frequency and damping rate have bên reduced due to the plus sign, in contrast to the minus sign of the Hamiltonian term in the total Hamiltonian describing the opto-mechanical coupling.} \label{fig_g2}
\end{figure}

In Fig. \ref{fig_g2}, contribution of the second order coupling is estimated. As we could see, the coupling $g_2$ could lead to a decrease in the modification of the effective mechanical frequency and the damping rate. This arises from the plus sign in the Hamiltonian term of second order process [Eq. (\ref{Htot})]. The second-order process could arise as a higher term in the Maclaurin series of the opto-mechanical interaction \cite{VyAPEX15} or naturally arise in a specific setting up of the system where the second-order process is excited, for example, the oscillator is inserted in the anti-node of an empty optical microcavity \cite{15Satya_quadratic,24ghorbani2024}. 

\section*{Acknowledgement}
We are thankful to Prof. Van-Hoang Le (HCMUE) for encouragement. This research is funded by Vietnam National Foundation for Science and Technology Development (NAFOSTED) under grant number 103.01-2023.36.

\bibliographystyle{cip-v3-bst-submit}
\bibliography{CiP_2024Vy}

\end{document}